
\input phyzzx.tex

\tolerance=1000

\twelvepoint

\normalbaselineskip=14pt


\REF\Julia{S.\ Ferrara, J.\ Scherk and B.\ Zumino, 
{\it Algebraic Properties of Extended Supersymmetry}, 
Nucl.\ Phys.\ {\bf B121} (1977) 393; 
E.\ Cremmer, J.\ Scherk and S.\ Ferrara, {\it SU(4) Invariant 
Supergravity Theory}, Phys.\ Lett.\ {\bf 74B} (1978) 61. B.\ Julia, {\it
Group Disintegrations}, 
in {\it Superspace \&
 Supergravity}, p.\ 331,  eds.\ S.W.\ Hawking  and M.\ Ro\v{c}ek, 
 E. Cremmer and B. Julia, {\it The $SO(8)$ Supergravity}, Nucl. Phys. {\bf
 B159} (1979) 141; B. Julia, {\it Infinite Lie Algebrans in Physics}, 
 Invited talk given at Johns Hopkins Workshop on Current Problems in 
 Particle Theory, Baltimore, Md., May 25-27, 1981.} 

\REF\Juliatwo{E. Cremmer and B. Julia,
{\it The $N=8$ supergravity theory. I. The Lagrangian},
Phys.\ Lett.\ {\bf 80B} (1978) 48.}

\REF\SW{J, Schwarz and P. West,
{\it Symmetries and Transformation of Chiral
$N=2$ $D=10$ Supergravity},
Phys. Lett. {\bf 126B} (1983) 301.}

\REF\Dual{E. Cremmer, B. Julia, H. Lu and C.N. Pope, 
{\it Dualisation of Dualities I}, Nucl. Phys. {\bf B535} (1998) 73, 
hep-th/9710119;  E. Cremmer, B. Julia, H. Lu and C.N. Pope, 
{\it Dualisation of Dualities II}, Nucl. Phys. {\bf B535} (1998) 242, 
hep-th/9806106.}

\REF\CJLP{E. Cremmer, B. Julia, H. L\"u and C.N. Pope, {\it
    Higher-Dimensional Origin of $D=3$ Coset Symmertries}, hep-th/9909099.}

\REF\peterone{P. West, {\it Hidden Superconformal Symmetries of M
theory}, JHEP, {\bf 0008} (2000) 007, hep-th/0005270.}

\REF\peter{P. West, {\it $E_{11}$ and M-Theory}, hep-th/0104081.}

\REF\rey{S-J, Rey, {\it Heterotic M(atrix) Strings and Their Interactions},
Nucl. Phys. {\bf B502} (1997) 170, hep-th/9704158.}

\REF\HS{G. Horowitz and L. Susskind, {\it Bosonic M-theory},
hep-th/0012037.}

\REF\MS{J. Maharana and J.H. Schwarz, {\it Noncompact Symmetries in
    String Theory}, hep-th/9207016.}

\REF\Ehlers{J. Ehlers, Dissertation, Hamburg University (1957)}

\REF\Geroch{R. Geroch, J. Math. Phys. {\bf 12} (1971) 918; 13 (1972) 394}
 
\REF\pope{C. Pope, {\it Lectures on Kaluza-Klein}, 
http://faculty.physics.tamu.edu/pope/.}

\REF\FT{E. Fradkin and A. Tseytlin, Nucl. Phys. {\bf 261} (1985) 1,
Phys, Lett. {\bf B155} (1985) 316.}

\REF\JN{B. Julia and H. Nicolai, {\it Conformal internal Symmetry of
    2d $\sigma$ models coupled to gravity and a dilaton}, 
Nucl. Phys. {\bf B482}
  (1996) 431, hep-th/9608082.}

\REF\Hermann{H. Nicolai, Phys. Lett. {\bf B187} (1987) 316.}

\REF\J{ B.\ Julia, {\it Group Disintegrations},
in {\it Superspace \&
Supergravity}, p.\ 331,  eds.\ S.W.\ Hawking  and M.\ Ro\v{c}ek,
Cambridge University Press (1981).}

\REF\Hermanntwo{H. Nicolai 
{\it A Hyperbolic Kac-Moody Algebra from Supergravity},
Phys. Lett. B276 (1992) 333. }

\REF\Maison{D. Maison, Phys. Rev. Lett. 41 (1978) 521; V. A. Belinskii and
V. E. Sakharov, Zh. Eksp. Teor. FIz. 75 (1978) 1955; 77 (1979) 3. }

\REF\Sol{{\it Solutions of Einstein's Equations: Techniques and Results}, 
ed C. Hoenselaers and W. Dietz; Springer-Verlag, Berlin (1984).}


\pubnum={KCL-TH-01-26\cr hep-th/0107209}

\date{July 2001}

\titlepage

\title{\bf Coset Symmetries in Dimensionally Reduced Bosonic String Theory}

\vskip 24pt

\centerline{N.D. Lambert and P.C. West\foot{lambert,pwest@mth.kcl.ac.uk}}

\address{Department of Mathematics\break
         King's College, London\break
         WC2R 2LS\break
         England\break
         }

\abstract

We discuss the dimensional reduction of various  effective actions, 
particularly that of the closed Bosonic string and pure gravity, 
to two and three dimensions. The result for the closed Bosonic 
string leads to coset symmetries which are in agreement with those 
recently predicted and argued to be present in a new unreduced
formulation of this theory. 
We also show that part of the Geroch group appears in the unreduced
duality symmetric  formulation of gravity recently proposed.
We conjecture that this formulation can be extended to  a
non-linear realisation based on a Kac-Moody algebra which we identify. 
We also briefly discuss the proposed action of Bosonic M-theory.

\endpage


\chapter{Introduction}

Early in the development of supergravity it was realised that 
the scalar fields belonged to a 
non-linear realisation [\Julia ]. 
Two of the most studied cases are the dimensional reduction
on a
$7$-torus of eleven-dimensional supergravity, which  leads to a maximal
supergravity whose scalars belong to $E_7$/SU(8) coset [\Juliatwo] and 
the IIB supergravity theory, whose scalars  belong to the coset
SU(1,1)/U(1) [\SW].  
The coset construction was extended [\Dual,\CJLP] to include the gauge
fields of supergravity theories. This method used generators that were
inert under Lorentz transformations and, as such, it is difficult to
extend this method to include either gravity or the Fermions. However,
this construction did include the  gauge and scalar fields as well as
their duals, and as a consequence  the  equations of motion for these
fields could be expressed as a generalised self-duality condition.

Recently the entire Bosonic sector, including gravity, 
of the eleven-dimensional and ten-dimensional IIA supergravity
theories  were formulated as non-linear realisations 
[\peterone]. Subsequently [\peter] it has been argued that 
eleven-dimensional supergravity and the ten-dimensional  IIA
supergravity  are invariant under a large rank eleven  Kac-Moody algebra
denoted $E_{11}$. This group includes the symmetries that  are found
when the theory is compactified on a
$n$-torus.  In particular it was conjectured  that M-theory is 
invariant under $E_{11}$. 
It was also proposed that  the effective action for the closed 
twenty-six-dimensional
Bosonic string possesses a rank twenty-seven  symmetry group  denoted 
$K_{27}$. In order to encode this symmetry it was shown that 
pure gravity could be 
reformulated in terms of two fields, which are related by duality 
transformation, and was also conjectured to possess an 
enlarged symmetry algebra. These proposals imply  a particular set of 
symmetries when these theories are dimensionally reduced on 
an $n$-torus. 

In this note we will systematically discuss the dimensional reduction
of the  effective action 
of the (non-supersymmetric) twenty-six-dimensional closed Bosonic string,
twenty-seven dimensional Bosonic M-theory (proposed in [\rey,\HS])
and pure gravity  and find the coset structure of the resulting scalar
degrees  of freedom. In the  case of the Bosonic string, 
reduced to three dimensions, we obtain the same $O(24,24)$
coset structure that was observed in [\CJLP] (also see [\MS] for
the compactification to four dimensions).
We will also  argue that further compactification to two dimensions
leads to a new group consistent with $K_{27}$.
Thus  our results agree with the symmetry groups    predicted in 
reference [\peter] and hence provide support for this proposal. For the 
case of pure gravity we will find the appearance of an enlarged
coset symmetry group in  the dimensional reduction to three dimensions
consistent with the old results of [\Ehlers, \Geroch]. We will show that
at least part of this symmetry is already present in the duality symmetric
formulation of gravity proposed in [\peter] before it is
dimensionally reduced. This encourages the hope that this formulation of
gravity possess symmetries that relate  many of the  solutions of general
relativity.  We will also suppose that the duality symmetric formulation
of gravity can be extended to a non-linear realisation based on a
Kac-Moody algebra. We conjecture that this is a rank $D$ Kac-Moody
algebra whose Dynkin diagram is given in figure 1:
$$
\matrix{
&&&0&&&&&&&&&\cr
&&&|&&&&&&&&&\cr
&&&0&&&&&&&&&\cr
&&&|&&&&&&&&&\cr
-&0&-&0&-&0&-&0&-&\ldots&
&-&0&-&0&-\cr
&&&&&&&&{\rm Fig.\ 1}\cr}
$$ 
where the first and last nodes on the bottom line are connected.

The rest of this paper is organised as follows. In section two
we discuss the dimensional reduction of various $D$-dimensional 
gravitational theories to two and three dimensions.
In particular we study the appearance of coset symmetries in
the compactified theory.
In section three we discuss how these symmetries arise in the
uncompactified using the formulation proposed in [\peter]. 
Finally we make some comments regarding the effective action
of Bosonic M-Theory.

In a sense our discussion in section two is 
complementary to that of [\CJLP], in that we
start with a family of higher dimensional actions and ask which ones lead
to a coset structure in three dimensions, whereas [\CJLP] starts
with a coset structure in three-dimensions and then lifts it
up to a higher dimensional theory. In addition we will be mainly
interested in non-supersymmetric actions. 
For the wide class of  theories we consider, we  will see that only
a very small set of actions lead to coset symmetries. In particular
we will find that M-theory type actions (i.e. gravity coupled to a 
four-form) only admit a coset
structure in the familiar case of eleven dimensions. 
On the other hand we will find string theory actions (i.e. gravity
coupled to a three-form and a dilaton) in all dimensions. 
However in this case the existence of a coset structure uniquely fixes the
coupling of the dilaton to the three-form. 
From the point of 
view of dimensional reduction the emergence of a coset structure is
somewhat miraculous. However one of the motivations of [\peter] was to
explain these cosets as symmetries in the uncompactified theory.
That these coset structures exist beyond the examples of supergravity
is further support for the programme of [\peter]. 


\chapter{Reducing Bosonic actions}

\section{Compactifications and Coset scalar Lagrangians}
Before starting with the reduction of gravitational theories it
will be helpful to first discuss some aspects of
compactifications and coset non-linear realisations. 
We will closely follow the method and construction outlined in [\pope].
Many of the technical steps required for the dimensional reductions
carried out in this section can also be found in [\CJLP]. However since
the aims of this paper are different, and in particular since we will
need to highlight certain points, in this  section we give  
a self contained account of the dimensional reduction procedures
and results.

As reviewed in [\pope], we choose to compactify so as to remain in
the `Einstein' frame with a standard kinetic term for the resulting
scalar. This fixes the compactification ansatz to be

$$
ds^2_{d+1} = e^{2\alpha_d\phi}ds^2_d 
+ e^{-4(d-2)\alpha_d\phi}(dx_{d}+A_\mu dx^\mu)^2\ ,
\eqn\compactification
$$
where 
$$
\alpha_d = \sqrt{1\over2(d-1)(d-2)}\ .
\eqn\sdef 
$$
Under such a compactification we find
$$
\int d^{d+1}x eR = \int d^dx  
e \left(R - {1\over2}\partial_\mu\phi\partial^\mu\phi
-{1\over 4}e^{-2(d-1) \alpha_d\phi}F_{\mu\nu}F^{\mu\nu}\right)\ ,
\eqn\Rredux
$$
where $F_{\mu\nu}=\partial_\mu A_\nu - \partial_\nu A_\mu$.
In addition will  also consider  $m$-form field strengths
$F_{\mu_1...\mu_m}$. Under the reduction 
\compactification\ we find
$$
\int d^{d+1}x {e\over 2 m!}F_{(m)}^2 = 
\int d^{d} x
{e\over 2 m!}e^{-2(m-1)\alpha_d\phi}F_{(m)}^2 
+{e\over 2 m!}e^{2(d-m)\alpha_d\phi}F_{(m-1)}^2\ .
\eqn\Fredux
$$
Note that it is understood in these equations that the metric
quantities are always those of the relevant dimension (e.g. in
\Fredux\ $e$ refers to the  $(d+1)$-dimensional volume element
on the left hand side and the
$d$-dimensional volume element on the right hand side).

Clearly we may repeat this proceedure to obtain a compactification
to on an $n$-torus.
In addition to the $n$ scalars $\phi_i$ that parameterise the
radii of the tori we will also find scalars from the $(m-1)$-form
gauge fields associated to $F_{(m)}$ along the compact directions. 

In addition, once we reach $m+1$ dimensions, we can dualise the $m$-form
field strengths into scalars. 
To do this we suppose that the action has the form
$$
\int d^{m+1} x {e\over 2 m!}e^{\vec\alpha\cdot\vec\phi}F_{(m)}^2  .
\eqn\dualone
$$
Next we no longer view
$F_{(m)}$ as the curl of a $(m-1)$-form but instead introduce  
a Lagrange multiplier term into the action
$$
\int d^{m+1} x {e\over 2 m!}e^{\vec\alpha\cdot\vec\phi}F_{(m)}^2 
+{e\over m!}\varphi\partial_\mu F_{\nu_1...\nu_m}\epsilon^{\mu\nu_1...\nu_m}\ .
\eqn\Lagrange
$$
Variation with respect to $\varphi$ simply enforces the Bianchi identity
on $F_{(m)}$. However we can also integrate by parts so that there are
no derivatives acting on $F_{(m)}$. Eliminating $F_{(m)}$ by its 
algebraic equation of motion 
$F^{\nu_1...\nu_m}=e^{-\vec\alpha\cdot\vec\phi}\epsilon^{\mu\nu_1...\nu_m}\partial_{\mu}\varphi$ and 
substituing back into \Lagrange\ leads to the equivalent scalar action
$$
\int d^{m+1}x {e\over 2}e^{-\vec\alpha\cdot\vec\phi}
\partial_\mu\varphi\partial^\mu\varphi\ . 
\eqn\dual
$$
Note the change in sign of the vector $\vec\alpha$.

Thus if we compactify all the way down to three dimensions
then we can dualise all fields into scalars and put the 
Lagrangian in the form
$$
{\cal L}
=e\left[R - {1\over2}\partial_\mu \vec{\phi}\cdot\partial^\mu \vec \phi
-{1\over2} \sum_{\vec \alpha}e^{\vec\alpha \cdot \vec\phi}
\partial_\mu\chi_{\vec\alpha}\partial^\mu\chi_{\vec\alpha}+\dots\right]\ .
\eqn\cosetL
$$
Here $\vec\phi = (\phi_1,...,\phi_n)$, the $\vec\alpha$
are constant $n$-vectors and $\chi_{\vec\alpha}$ are additional scalar
fields labelled by $\vec\alpha$. We see that there are effectively
two type of scalars, those labelled by $\vec\phi$ and those
labelled by $\chi_{\vec\alpha}$. The distinguishing feature between
these two types of scalars is that various gauge symmetries imply
that the $\chi_{\vec\alpha}$ scalars can only have derivative interactions.
By definition the $\vec\phi$ scalars have non-derivative interactions
and  are obtained from the diagonal components of the metric and we will
see that the may be associated to the Cartan subalgebra when a coset
structure exists.

In fact our discussion and  \cosetL\ are somewhat 
over simplified since the off-diagonal metric components (i.e.
components of the graviphotons) affect the dualisation argument
by altering the Bianchi identity for $F_{(m)}$ in \Lagrange. 
In addition there maybe Chern-Simons terms in the uncompactified
theory which will also alter the constraint in \Lagrange.  
The effect of these modifications is to 
replace the kinetic terms $\partial_\mu \chi_{\vec\alpha}$ by 
$ A_{\vec\alpha}^{\ \vec\beta}\partial_\mu\chi_{\vec\beta}$, 
where $A_{\vec\alpha}^{\ \vec\beta}$ is a non-degenerate
matrix depending on
the scalars $\chi_{\vec\alpha}$. However we may safely ignore these
subtlties here. Their presence is denoted by the ellipsis in
\cosetL.

On the other hand we wish to identify \cosetL\ with a $G/H$
non-linear realisation. To this end we consider a group $G$ with 
generators ${\vec H}$ and $E^{\vec\alpha}$  where the Cartan
subalgebra is generated by 
$\vec H$ and 
$$
[\vec H, E^{\vec \alpha}] = \vec \alpha E^{\vec \alpha}\ .
\eqn\algebra
$$
We remind the reader that we may split the roots into positive and
negative ones. In what follows we call a root  positive (negative)
if it's first non-vanishing element, as counted from the right,
is positive (negative). 
In addition the positive roots can be written as linear combinations,
with non-negative integer coefficients, of the so-called simple roots.
The Cartan matrix is defined in terms of the simple roots to be
$$
C_{ij}=2{\vec\alpha_i\cdot\vec\alpha_j\over\vec\alpha_i\cdot\vec\alpha_i}\ .
\eqn\cartan
$$

Provided that the roots  obey the Serre relation, the form of the
Cartan matrix uniquely determines the  algebra of $G$.
It will be important later to recall that a defining property of
a Kac-Moody algebra is that the off-diagonal entries 
of the Cartan matrix are negative integers or zero.
Every Kac-Moody algebra admits a so-called Cartan involution
$\tau:(E^{\vec\alpha},E^{-\vec\alpha},\vec H)
\rightarrow -(E^{-\vec\alpha},E^{\vec\alpha},\vec H)$. The Cartan
involution can be used to define a subgroup, namely, the subgroup which
is invariant under it.

For simplicity we consider non-linear realisations for which
the local subalgebras of the  Kac-Moody algebra are chosen to be those
which are invariant under the Cartan involution. In the case that the
subgroup is maximally noncompact the coset representatives can be written
as
$$
{\cal U} = e^{{1\over 2}\vec \phi \cdot \vec H}
\sum_{\vec \alpha >0}e^{\chi_{\vec \alpha}\cdot E^{\vec\alpha}}\ ,
\eqn\Udef
$$
where the sum is only over the positive  roots.
It is not hard to show that the scalar Lagrangian
$$
{\cal L}_{G/H} = {1\over 4}{\rm Tr}\left(
\partial_\mu{\cal M}^{-1}\partial^\mu{\cal M}\right)\ ,
\eqn\Ldef
$$
where
$$
{\cal M} = {\cal U}^\#{\cal U}\ ,
\eqn\Mdef
$$
is invariant under global $G$ transformations and local $H$
transformations, ${\cal V}\rightarrow h(x){\cal V}g$. 
Here we have used a generalised transpose acing on the generators: 
$X^\#=\tau(X^{-1})$.
In the simplest cases, corresponding to orthogonal subgroups $H$, 
$\#$ coincides with the  transpose.
If we normalise the Cartan and positive root generators so that 
$$
{\rm  Tr}(H_iH_j)=2\delta_{ij}\ ,\qquad 
{\rm  Tr}(E^{\vec\alpha} E^{\vec\beta})=0\ ,\quad
{\rm  Tr}\left( {E^{\vec\alpha}}
E^{-\vec\beta}\right)=\delta^{\vec\alpha\vec\beta}\ ,
\eqn\norm
$$
then one can show that ${\cal L}_{G/H}$ is precisely the scalar part
of \cosetL. Thus it follows that if the vectors $\vec\alpha$
obtained from the compactification can be identified with the
positive roots of a group $G$, then the action when dimensionally
reduced to three dimensions  has a $G/H$ symmetry.

\section{Pure Gravity}

It is instructive to first consider the example of pure gravity
in $D$ spacetime dimensions:
$$
S = \int d^{D}x e R\ .
\eqn\gravity
$$
One class of scalars arises by starting with the curvature
scalar in $D-k$ dimensions, then reducing to a 2-form field
strength in $D-k-1$ dimensions and then reducing this to a
scalar in $D-k-2$ dimensions. In a sense this is the
`fastest' way to obtain a $\chi$-scalar starting with the metric
in $D-k$ dimensions. 

Reducing pure gravity in $D$ dimensions on an $n$-torus results
in $n-2$ such scalars with the roots
$$
\vec\alpha_k =
(0,...,0,-2(D-k-2)\alpha_{D-k-1},2(D-k-4)\alpha_{D-k-2},0,...,0) \ ,
\eqn\alphadef
$$
where $\alpha_{d}$ is given in \sdef, 
$k=0,...,n-2$ and there are $k$ zeros on the left
and $n-2-k$ zeros on the right.
It is not hard to see that these satisfy
$$
\vec\alpha_k\cdot\vec\alpha_l = \left\{\matrix{
4\cr-2\cr 0\cr} \ 
\matrix{k=l\cr|k-l|=1\cr|k-l|\ge2\cr}\right.
\eqn\alphaprop
$$

Other methods of obtaining 
$\chi$-scalars, i.e. by not going directly from a two-form to
a scalar, result in an additional 
${1\over2}(n-1)(n-2)$ scalars. However their associated  roots 
are not simple but instead take the form
$\vec\alpha_k+\vec\alpha_{k+1}+...+\vec\alpha_{k+l}$ where
$k=0,...,n-2$ and $l=1,...,n-k-2$.
These roots are summarised in the Dynkin diagram $A_{n-1}$ 
associated to the group $SL(n)$
$$\matrix{
0&-&0&-&\ldots&-&0&-&0\cr
&&&&{\rm Fig.\ 2}\cr}
$$
for $n\le D-4$.

Now in three dimensions (i.e. for $n=D-3$) 
we can use \dual\ to dualise the graviphotons
into scalars. Dualising the first
vector that arises (i.e. after compactifaction to $(D-1)$ dimensions 
on a single $S^1$) leads to the root
$$
\vec\delta = (2(D-2)\alpha_{D-1},2\alpha_{D-2},2\alpha_{D-3},
\ldots,2\alpha_3)\ .
\eqn\deltadef
$$
This root is positive and,  since its first entry is positive 
whereas none of the
$\vec\alpha_k$ have a positive entry in the first column, it is also simple.
One can show that 
$$
\vec\delta\cdot\vec\alpha_k = \left\{\matrix{
0\ \ \ k\ne  0\cr
-2\ \ \ k= 0\cr}\right. \ ,\quad \vec\delta\cdot\vec\delta =4\ .
\eqn\deltaprop
$$

On the other hand dualising the remaining vectors in three-dimensions
leads to $D-4$ non-simple positive roots of the form 
$\vec\delta+\vec\alpha_0+\vec\alpha_1+...+\vec\alpha_l$, $l=0,...,D-3$.
Thus find that the action of gravity reduced to three-dimensions
has a coset structure and the associated  Dynkin diagram $A_{D-3}$
$$\matrix{
0&-&0&-&0&-&\ldots&-&0&-&0&-&0\cr
&&&&&&{\rm Fig.\ 3}\cr}
$$
corresponding to the group $SL(D-2)$.
Note that two simple roots, the first and last in the diagram,  
appear upon compactification to three-dimensions.
Thus the Cartan subalgebra is $D-3$-dimensional and there 
are a total of ${1\over2}(D-2)(D-3)$ positive roots, with $D-3$ simple roots.

\section{Gravity Coupled to a Three-Form}

Next we wish to consider the compactifications of actions which
include a dilaton $\phi$ and a three-form field strength
$H_{\mu\nu\rho}$. 
In particular we consider  actions of the form
$$
S= \int d^{D}x e\left(R 
-{1\over2} \partial_\mu\phi\partial^\mu \phi
-{1\over 12}e^{\beta\phi}H_{\mu\nu\lambda}H^{\mu\nu\lambda}\right)\ ,
\eqn\stringact
$$
where $\beta$ is a constant.
For a special choice of $\beta=\sqrt{8/D-2}$, \stringact\ is
the effective action of the Bosonic string (with the tachyon 
set to zero) or the NS-NS sector of the superstring. 
The reduction of \stringact\  was 
previously discussed in [\MS]. However the possibility of
dualising the vectors in three dimensions was not performed and hence
the corresponding coset was reduced.

Clearly  if we dimensionally reduce this system on an $n$-torus
we find the same scalars and roots that we did for pure
gravity, coming from the $eR$ term in \stringact. 
For this action we also begin with an 
additional scalar $\phi$ already in $D$ dimensions.
The existence of $\phi$ implies  that
the putative root vectors are now $(n+1)$-dimensional. 
In particular the roots $\vec\alpha_k$ discussed in 
section 2.2 gain an extra column on their left containing a zero.
We will also obtain additional putative roots from dimensional
reduction of the three-form.

Let us now consider the scalars and their roots that originate from
the three-form.
The first scalar from $H_{\mu\nu\rho}$ arises when $n=2$ this gives
a root
$$
\vec\beta = (\beta,2(D-4)\alpha_{D-1},2(D-4)\alpha_{D-2},0,...,0)
\eqn\betaroot
$$
where there  are $n-2$ zeros on the right. A little work shows that
$$
\vec\beta\cdot\vec\alpha_k = \left\{\matrix{0\cr-2\cr}\matrix{k\ne 1\cr
k=1\cr}\right.\ ,\quad \vec\beta\cdot\vec\beta= \beta^2+4{D-4\over D-2}
\ .
\eqn\betaprop
$$

Next we need to ensure that the Cartan matrix elements \cartan\ are
non-positive integers. For $D>6$ the only possibility is to fix 
$\vec\beta\cdot\vec\beta=4$, i.e. $\vec\beta$ is the same length as
the $\vec\alpha_k$ roots. This in turn determines the coupling to the
three-form to be
$$
\beta=\sqrt{8\over D-2}\ .
\eqn\alphazero
$$
This dilaton coupling $\beta$ is precisely that
found in the Bosonic string effective action in $D$ dimensions
[\FT].

Continuing with the compactification on an $n$-torus, with $n\le D-5$,
we find additional scalars and
their (positive) roots. However they are not simple but rather of
the form $\vec\beta+\sum_k n_k\vec\alpha_k$ 
where $n_k$ are positive integers.
These roots are summarised by the Dynkin diagram $D_{n}$
$$
\matrix{
&&&&&&&&0&\cr
&&&&&&&&|&\cr
0&-&0&-&\ldots&-&0&-&0&-&0\cr
&&&&&{\rm Fig.\ 4}\cr}
$$
whose maximally non-compact form is $O(n,n)$.

Upon compactification to 
four dimensions we may dualise $H_{\mu\nu\lambda}$ 
to obtain a scalar and its associated root 
$$
\vec\gamma= (-\beta, 4\alpha_{D-1},4\alpha_{D-2} ,...,4\alpha_4)\ ,
\eqn\gammaroot
$$
One can explicitly check that $\vec\gamma$ is simple and satisfies
$$
\vec\gamma\cdot\vec\alpha_k=0\ , \quad
\vec\gamma\cdot\vec\beta=0\ ,\quad 
\vec\gamma\cdot\vec\gamma=\beta^2+4{D-4\over D-2}\ ,
\eqn\newgammaprop
$$
where $k=0,...,D-6$. Note that $\vec\gamma$ 
automatically has the same length as
$\vec\beta$.

Continuing to three dimensions we obtain more
putative roots
by dualising the vectors. However, in contrast to section 2.2 we find
that none of these give simple roots. In particular we now find that
$\vec\delta=\vec\beta+\vec\gamma+\vec\alpha_0+...+\vec\alpha_{D-5}$
and hence $\vec\delta$ is no longer simple. In addition we find that 
$\vec\gamma\cdot\vec\alpha_{D-5}=-2$. Thus the simple
roots are those of the
Dynkin diagram $D_{D-2}$ and corresponding maximally non-compact group 
$O(D-2,D-2)$
$$
\matrix{
0&&&&&&&&&&0&\cr
|&&&&&&&&&&|&\cr
0&-&0&-&0&-&\ldots&-&0&-&0&-&0\cr
&&&&&&{\rm Fig.\ 5}\cr}
$$

Here the roots on the bottom row are the roots $\vec\alpha_k$ 
from section 2.2 and appear from right to left as we compactify,
with the first (right most) simple root
arising in $D-2$ dimensions 
and the last (left most) simple root arising in three
dimensions. In the top row the right root $\vec\beta$ appears after
compactification to $D-2$ dimensions and the left root $\vec\gamma$ 
appears upon compactification to four dimensions.

We conclude this section  by noting that for $D\le 6$ there is another
possible choice for $\beta$ that leads to an integral Cartan matrix.
Namely we could set $\beta = \sqrt{12-2D/D-2}$ so that the  
length of $\vec\beta$ and $\vec\gamma$ is $\sqrt{2}$. However in this
case we find that $\vec\beta\cdot\vec\gamma=2$, which is positive and
hence can not be identified with a Kac-Moody algebra. 

\section{Gravity Coupled to a Four-Form}

Finally  we can consider  gravity coupled to a
four-form $G_{\mu\nu\lambda\rho}$ with no scalar in $D$ dimensions. 
This occurs in M-theory in eleven dimensions and
more recently it was conjectured to be the effective action for a
Bosonic M-theory in twenty-seven
dimensions [\HS]. In this case we start with the action
$$
S= \int d^{D}x e\left(R 
-{1\over 48}G_{\mu\nu\lambda\rho}G^{\mu\nu\lambda\rho}\right)\ .
\eqn\Mact
$$
Clearly when we dimensionally reduce this action on an $n$-torus we obtain
all of the roots we found in section 2.2. In addition we find
scalars and their roots from compactifying the four-form. 
The first such root arises for $n=3$ and is
$$
\vec\epsilon = 
(2(D-5)\alpha_{D-1},2(D-5)\alpha_{D-2},2(D-5)\alpha_{D-3},0,...,0)\ ,
\eqn\epsilonroot
$$
where there are $n-3$ zeros on the right.
It is straightforward to check that
$$
\vec\epsilon\cdot\vec\alpha_k= \left\{\matrix{0\cr-2\cr}\matrix{k\ne 2\cr
k=2\cr}\right.\ ,
\quad \vec\epsilon\cdot\vec\epsilon=6{D-5\over D-2}\ .
\eqn\epsilonprop
$$
Further compactification to lower dimensions introduces additional
scalars and their roots. However one can check that none of these
are simple. 

In addition, in five dimensions, we can dualise
$G_{\mu\nu\lambda\rho}$ to obtain a new root $\vec\eta$, with the same 
length as $\vec\epsilon$. While $\vec\eta$
is always a positive linear combination of $\vec\epsilon$ and 
$\vec\alpha_k$,
the coefficients are integers only if $D=3a+5$ for a non-negative 
integer $a$. Similarly in 
three-dimensions $\vec\delta$ is a positive linear combination of the 
other roots but only with  
integer coefficients if $D=3b+2$ for a non-negative integer $b$.

Thus only eleven-dimensional M-theory has 
$\vec\epsilon\cdot\vec\epsilon=4$ and we find the Dynkin
diagram $E_8$
(as first observed in [\Julia])
$$\matrix{
&&&&&&&&0&&&&&\cr
&&&&&&&&|&&&&&\cr
0&-&0&-&0&-&0&-&0&-&0&-&0\cr
&&&&&&{\rm Fig.\ 6}\cr}
$$
As above the bottom row of roots $\vec\alpha_k$ are those of section
2.2 and appear  from right to left as we compactify, 
starting with the reduction to nine dimensions on the right and
ending in three dimensions on the left. The root $\vec\epsilon$ on
the top row appears in eight dimensions.
For any other dimension
the Cartan metric \cartan\ does not have integer entries.  
Hence only eleven-dimensional M-theory possesses a coset symmetry.

We may also consider the effective action of the type II and type 0 
strings. Of course it is well-known that the effective action of 
type IIA and type IIB string theories, when reduced on an $n$-torus, is
identical to the effective action of M-theory reduced on an
$n+1$-torus.  Therefore these theories also have the same
coset symmetry as eleven-dimensional M-theory. The type 0A and type
0B string theories differ from the type IIA and type IIB theories
in that they do not have any Fermions but instead have double the number 
of Ramond-Ramond fields. These theories are identical
to each other after compactification on an $n$-torus but differ
from eleven-dimensional M-theory. After
compactification to three-dimensions they clearly contain all the
roots that one finds from compactification of eleven-dimensional
M-theory. However there will also be additional roots from the doubled
Ramond-Ramond fields. In fact these will just provide a doubling of
some roots, but not all, in the compactification of
M-theory. Therefore, although one can identify a root system, it is not clear
that the action can be written as a non-linear realisation.

\section{Further Reduction to Two Dimensions}

In the previous three sections we discussed the toroidal
compactification of pure gravity, string theory and M-theory 
from $D$ dimensions to three dimensions. This leads to
an action of the form \cosetL, consisting only of scalars
coupled to gravity. However in three dimensions and less gravity
has no degrees of freedom. Thus further compactification will
not lead to new scalars in the coset Lagrangian. 
On the other hand the symmetry of the system is increased upon
compactification to two dimensions. In particular, as described in
[\JN], the global $G$ symmetry of the coset 
action \cosetL, when reduced to two dimensions, can be enlarged to
an affine symmetry. Thus the total symmetry of group is expected
to be an affine version of the groups $G$ found above, obtained
by adding a single root in two dimensions to the Dynkin diagrams
above. It is natural to assume  that this new root $\vec\zeta$
has length two and 
is perpendicular to all the other roots except
$\vec\alpha_{D-5}$ and $\vec\delta$ which arise in three-dimensions.
For these roots we assume that
$\vec\zeta\cdot\vec\alpha_{D-5}=\vec\zeta\cdot\vec\delta=-2$. 
As a check on this assumption we note that the Dynkin diagrams 
we obtain in two dimensions for pure gravity and eleven-dimemsional 
M-theory are
$$\matrix{
-&0&-&0&-&0&-&0&-&\ldots&-&0&-&0&-&0&-\cr
&&&&&&&&{\rm Fig.\ 7}\cr}
$$
where the first and last roots are connected, and 
$$
\matrix{
&&&&&&&&&&0&&&&&\cr
&&&&&&&&&&|&&&&&\cr
0&-&0&-&0&-&0&-&0&-&0&-&0&-&0\cr
&&&&&&&{\rm Fig.\ 8}\cr}
$$
respectively. Thus we obtain the Dynkin diagrams of affine $SL(D-2)$
and $E_9$ for pure gravity and eleven-dimensional M-theory respectively.
These diagrams are those of an affine algebra and are consistent with the
results of [\Ehlers, \Geroch] for four dimensional gravity and agree with
the results of [\Hermann,\J] for M theory. 

Returning to the effective action of the Bosonic string, which is our
main subject here,  
we find the Dynkin diagram
$$
\matrix{
&&0&&&&&&&&&&0\cr
&&|&&&&&&&&&&|\cr
0&-&0&-&0&-&0&-&\ldots&-&0&-&0&-&0\cr
&&&&&&&{\rm Fig.\ 9}\cr}
$$
As a further check we note that this is indeed an affine algebra.


\chapter{Symmetries of the Uncompactified Theory}

\section{The Bosonic String}

In [\peter] it was argued that effective action for the closed
Bosonic string is invariant under a Kac-Moody algebra of rank 27, 
denoted by $K_{27}$ and whose Dynkin diagram is
$$
\matrix{
&&&&&&0&&&&&&&&&0\cr
&&&&&&|&&&&&&&&&|\cr
0&-&0&-&0&-&0&-&0&-&\ldots&
&-&0&-&0&-&0\cr
&&&&&&&&{\rm Fig.\ 10}\cr}
$$ 
In this section we will  show that the symmetries found earlier 
in the dimensionally reduced effective  action of the
closed Bosonic string (here we set $D=26$) 
are  indeed those predicted in [\peter].  Let us first recall the
derivation of the expected symmetry from the
$K_{27}$ algebra. We consider the restriction of the $K_{27}$ algebra
obtained by taking all the indices on the generators to lie between 
$26$ and $26-n+1$. In terms of the Dynkin diagram of
$K_{27}$ in figure 10 this is equivalent to  keeping only the first $n-1$
nodes on the horizontal line of the Dynkin diagram starting from the
right, as well as any nodes to which they are attached by vertical
lines.  The Dynkin diagram so
obtained is $D_{n}$ provided $3\le n \le 22$, $D_{24}$ if $n=23$
and affine $D_{24}$  if $n=24$. 
It is natural to take the
appropriate real form to be  $O(n,n)$ for  $3\le n \le 22$ and 
$O(24,24)$ if $n=23$. The corresponding local subgroups are  taken to
be those  left invariant under the Cartan involution. For these
groups, the local subgroups are then  
$O(n)\times O(n)$ for $3\le n \le 22$ and 
$O(24)\times O(24)$ if $n=23$.

The symmetry groups found in the above restriction of the $K_{27}$ algebra
are part of the proposed symmetries of the closed Bosonic string in 
twenty-six dimensions. However, they are also the symmetries that we expect to
find in the dimensional reduction on a $n$-torus. One way to see this is 
to note that dimensional reduction on a $n$-torus leads to a $SL(n-1)$
symmetry resulting from general coordinate transformations that are
preserved by the reduction ansatz. This $SL(n-1)$ symmetry is just the 
subalgebra of the Dynkin diagram of $K_{27}$ which results from 
deleting all nodes in the diagram except the first $n-1$ horizontal
nodes. The identification is confirmed by checking that the group 
action of $SL(n-1)$ on the coordinates, as derived from the
non-linear realisation, agrees with the general
coordinate transformations which are preserved in the reduction
procedure.  

For $n\le 22$ one expects to find that the scalars belong to the coset 
$O(n,n)/O(n)\times O(n)$. In fact the reduction of the effective action
of  the closed Bosonic string was carried out some time ago in [\MS] 
for a $n$-torus with $n\le 22$ and these authors did find this
coset.  It was noted in [\MS] that in three dimensions one had the
possibility to dualise the one form gauge fields and this might change
the resulting coset symmetry. 
In section 2.3 above,   upon dualising the vectors in three dimensions, we 
founded that the scalars belonged to the coset 
$O(24,24)/O(24)\times O(24)$ in agreement with [\CJLP] and as predicted 
by the $K_{27}$ algebra of the twenty-six-dimensional theory.
This provides an important  check on the proposed $K_{27}$ symmetry of the
closed Bosonic string.  

A further check is provided by the reduction to two dimensions. 
As discussed above the $K_{27}$ symmetry of the effective action of
the twenty-six-dimensional  Bosonic string predicts that the symmetry
that occurs in the reduction is that found by keeping  23 of
the horizontal nodes of the Dynkin diagram and any  vertical lines
attached to these. Examining figure 10 we see that the symmetry is
{\it affine} $O(24,24)$. This is precisely the symmetry that occurs in the
dimensional reduction as explained in section 2.5. 

\section{Pure Gravity}

Let us now turn to the analogous discussion
of pure gravity. In [\peter] it was found that in order to exhibit the
proposed $E_{11}$ invariance of eleven-dimensional supergravity, 
its gravitational degrees of freedom would have to be
described by a formulation which  (when generalised to
$D$ dimensions)  includes  
a  dual field $h_{a_1\ldots a_{D-3},b}$
as well as the usual  $h_a{}^b$ field. It was
suggested that this formulation of gravity could be formulated as a
non-linear realisation based on an algebra that included 
the generators
$K^a{}_b$ of $SL(n-1)$ and in addition 
generators $R^{a_1\ldots a_{D-3},b}$ associated to $h_{a_1\ldots a_{D-3},b}$.
These additional generators
are antisymmetric in their $a_1\ldots a_{D-3}$ indices and satisfy 
$R^{[a_1\ldots a_{D-3},b]}=0$. They obey the commutators
$$\eqalign{
[K^a{}_b,\ K^c{}_d]&=\delta_b^c K^a{}_d-\delta^a_d K^c{}_b\ ,\cr 
[ K^c{}_d,\ R^{a_1\ldots a_{D-3},b}]&= \delta_d^{a_1}R^{c a_2\ldots
a_{D-3},b}+\ldots +\delta_d^b R^{a_1\ldots a_{D-3},c}\ ,\cr}
\eqn\ptwo
$$
where the ellipsis denotes the appropriate antisymmetrisation. 

As above we now consider the subalgebra that is found  in
the dimensional reduction of pure gravity on an $n$-torus. 
This subalgebra is found
by only  considering the generators whose indices lie between
$D$ and
$D-n+1$. If $n < D-3$  this subalgebra only consists 
the  generators $K^i{}_j,\ i,j=D,\ldots ,D-n+1$ of 
$GL(n)$. However if $n=D-3$ then we must also include the generator 
$R^{i_1\ldots i_{D-3},j}$. This latter generator may be written as 
$$
R^{i_1\ldots i_{D-3},j}=\epsilon^{i_1\ldots i_{D-3}} S^j\ .
\eqn\pppp
$$ 
It is straightforward to show that the elements
$$\eqalign{
\hat K^i{}_j &= K^i{}_j - {1\over (D-3)}\delta_j^i \sum_k K^k{}_k\ , \cr
\hat K^2{}_j&= S^j\ ,\cr
\hat D&=\sum_k K^k{}_k\ ,\cr }
\eqn\pthree
$$ 
generate $GL(D-2)$, except for the generator $\hat K^j{}_2$. 
The absence of this generator can be accounted for by the fact that the
algebra given in \ptwo\ is only a part of the underlying
symmetry of this dual formulation of gravity. This is  much 
the same as the way that $G_{11}$ is
only part of the proposed $E_{11}$ symmetry of eleven-dimensional
supergravity. 
If one started from the full underlying symmetry one would expect to 
find the full $GL(D-2)$ algebra.  We return to this point shortly. 

As discussed in section 2.2 and 2.3  these are indeed the coset
symmetries that result when pure gravity is dimensionally reduced. This
includes the enlargement of the symmetry in three
dimensions due to the dualisation of the vector fields. This symmetry
gives rise to
an $SL(2,{\bf R})$ symmetry discovered long ago  [\Ehlers, \Geroch] in
the reduction of four dimensional gravity. As anticipated in [\peter]
we have shown here that the Borel subgroup of  this symmetry is part of
the symmetry of the new formulation of gravity proposed in [\peter],
even before it is dimensionally reduced. It is known [\Geroch,\JN, \J,
\Hermanntwo,\Maison] that this $SL(2,R)$  symmetry becomes 
affinised if one continues the reduction to two dimensions. These 
results have proved useful in the construction of solutions of
Einstein equations as they rotate one solution into another, for a review
see [\Sol]. We  expect that this affinised symmetry and the
symmetry that results from the reduction to one dimension   is also part
of the new formulation of gravity given in [\peter]. 
This suggests that in this dually symmetric formulation one may find that 
many solutions of Einstein's theory are in fact related by the symmetries
of  this formulation. 

At first sight this may seem to be a paradox since a symmetry in one
formulation of a theory is usually a symmetry in an equivalent formulation. 
However,  one usually
takes a symmetry to mean a local symmetry and what may appear as a local
symmetry when expressed in terms of the metric  may not be  a local
symmetry when expressed in terms of the dual field. By using both
variables in the dual formulation of gravity of [\peter]  one can then
find a larger set of local symmetries than appear in Einstein's
formulation of gravity. 

Although the 
generators $K^a{}_b$ and $R^{a_1\ldots a_{D-3},b}$  can be used
to construct a non-linear realisation describing  gravity, one
might hope to find a non-linear realisation based on a Kac-Moody algebra
as was proposed  for the maximal supergravity theories and the twenty-six
dimensional effective action of the Bosonic string. As in these cases
[\peter], one may try to identify the Kac-Moody algebra by finding the
simple positive roots and the Cartan subalgebra generators. 
Among the generators $K^a{}_b$ and $R^{a_1\ldots a_{D-3},b}$   we
recognise a set of $D$ commuting generators given by 
$ K^a{}_a$. The remaining generators, apart from the negative root
generators of $SL(D)$  can be found from multiple commutators of the $D$
generators 
$$
K^a{}_{a+1},\ R^{4\ldots D,D}\ , 
\eqn\Kgens
$$ 
which we may identify with the positive simple roots of the 
rank $D$ Kac-Moody algebra we are searching for.  As explained in [\peter]
this procedure is not unambiguous without the negative simple roots. 
However, the ambiguity may be resolved by identifying  
the Borel subalgebras of some particular subalgebras. 

Clearly the $SL(D)$ Borel
subalgebra has the simple positive roots 
$$
E_a=K^a{}_{a+1},\ a=1,\ldots, D-1\ ,
\eqn\pten
$$ 
and the Cartan subalgebra 
$$ 
H_a= K^a{}_{a}-K^{a+1}{}_{a+1}\ , \ a=1,\ldots, D-1\ .
\eqn\peleven
$$ 

The other
Borel subalgebra  we might like to identify is affine $SL(D-2)$ or
$A^{(1)}_{D-3}$. However, the generators considered above do
not contain all of this Borel subalgebra. Nonetheless, it is tempting to
identify its simple roots as 
$$
E_a\ ,\ a=3,\ldots ,D-1\ ,
\eqn\po
$$
and
$$ 
E_D=R^{4\ldots D,D}\ .
\eqn\ptwelve
$$ 
The Cartan subalgebra elements are identified as 
$$
H_a\ , \ a=3,\ldots D-1\ ,
\eqn\pfourteen
$$
and 
$$ 
H_D=K^4{}_4+\ldots
+K^{D-1}{}_{D-1}-{D-6\over D-2}\sum_aK^a{}_a\ .
\eqn\pthirteen
$$
One can verify that these
do indeed satisfy the relation 
$[ H_a, E_b]=A_{ab}E_b$ where $A_{ab}$ is the Cartan matrix of
$A^{(1)}_{D-3}$. 

If we assume that these two subgroups have been correctly identified
then  the simple roots of the Kac-Moody algebra underlying the dual
formulation of gravity are $E_a,\ a=1,\ldots ,D$  and
Cartan subalgebra is $H_a,\ a=1,\ldots ,D$.  One then  finds that the Cartan
matrix resulting from their commutators corresponds to the Dynkin diagram
in figure 1.

\chapter{Bosonic M-Theory}

Finally we would like to briefly comment on the effective action of
``Bosonic M-theory''. It was proposed  in [\rey,\HS] that the strong 
coupling limit of the Bosonic string should be described by a 
twenty-seven-dimensional theory, called Bosonic M-theory. 
In particular the  authors of [\HS] proposed a twenty-seven-dimensional 
action of the form \Mact. However we have seen that this action  
doesn't have any coset symmetries when it is dimensionally reduced. On
the other hand it is supposed to represent a strong coupling limit of the
Bosonic string which, as we have seen, does have a large coset symmetry. 
It was noted in [\HS] that the compactification of Bosonic M-theory
to the Bosonic string does not quite give the correct action,
in particular the dilaton kinetic term is off by a factor of
$125/121$. This is effectively the same as the obstruction 
to obtaining a coset symmetry that we encountered in section 2.4. 
It was argued in [\HS] that, since there is no supersymmetry, 
the reduced Bosonic M-Theory action which is valid at strong coupling
does not need to agree numerically  with the
effective action of perturbative Bosonic string theory. Indeed without
supersymmetry or some other symmetry one  would not  expect it to. 
However from the
point of view taken here these coefficients are fixed by the coset symmetry, 
which includes T-duality, and one expects that 
the effective action of Bosonic M-theory should  have
all the symmetries of the Bosonic string, namely $K_{27}$.

One might try to invoke a different action for Bosonic M-theory.
For example we could consider a three-form instead of a four-form. 
In this case we 
encounter the same problem since, as we saw in section 2.3, 
demanding that the root $\vec\beta$ has the correct length is related to
the coupling of the  dilaton. Therefore proceeding
in this manner requires that we also add a dilaton. 
Indeed with this approach we simply end up with the effective action of a 
twenty-seven dimensional string theory. Alternatively
we could add a  dilaton into the Bosonic
action of [\HS]. However in this case one finds that there is no value for
the coupling $\beta$ between the dilaton and the four-form 
which leads to a Cartan matrix with integer entries.
Therefore if we assume that  a twenty-seven-dimensional
Bosonic M-theory  exists and reduces to give a coset
structure, then we must conclude that it doesn't have a simple, local
low energy effective action of the form we are familiar with.


\chapter{Acknowledgements}

This work was supported in part by the two EU networks entitled  "On 
Integrability, Nonperturbative effects, and Symmetry in Quantum Field 
Theory"  (FMRX-CT96-0012) and "Superstrings" (HPRN-CT-2000-00122). 
It was also  supported by the PPARC special grant PPA/G/S/1998/0061
and N.D.L. is support by a PPARC fellowship. One of the authors (PAW)
would like to thank L. Mason and G. Segal for discussions.

\refout

\end